\ifx\synctex\undefined\else\synctex=1\fi
\pdfoutput=1
\documentclass{eptcs}

\providecommand{\event}{VPT 2015}

\usepackage{graphicx}
\usepackage{amsmath} 
\usepackage{algorithm2e}
\usepackage{hyperref}
\usepackage{amssymb}
\usepackage{upquote}
\usepackage{fancyvrb}

\usepackage{wrapfig}
 
\usepackage[draft]{commenting}

\declareauthor{jg}{John}{blue}
\declareauthor{pg}{Pierre}{red!50}
\declareauthor{bk}{Bisho}{orange!50}

\title{ Solving non-linear Horn clauses using a linear solver}
\author{Bishoksan Kafle \institute{Roskilde University, Denmark} \email{kafle@ruc.dk}
 }

\def\titlerunning{Solving non-linear Horn clauses using a linear solver}
\def\authorrunning{ Kafle}

\begin{document}
\maketitle

\newcommand{\integ}{{\sf int}}
\newcommand{\listint}{{\sf listint}}
\newcommand{\other}{{\sf other}}
\newcommand{\true}{\mathsf {true}}
\newcommand{\false}{\mathsf {false}}
\newcommand{\Bin}{{\sf Bin}}
\newcommand{\Dep}{{\sf Dep}}
\newcommand{\g}{{\sf g}}
\newcommand{\nong}{{\sf ng}}
\newcommand{\OL}{{\cal O}}
\newcommand{\M}{{\sf M}}
\newcommand{\R}{{\cal R}}
\newcommand{\A}{\mathcal{A}}

\newcommand{\body}{\mathcal{B}}
\newcommand{\B}{{\cal B}}
\newcommand{\C}{{\cal C}}
\newcommand{\D}{{\cal D}}
\newcommand{\X}{{\cal X}}
\newcommand{\V}{{\cal V}}
\newcommand{\Q}{{\cal Q}}
\newcommand{\F}{{\sf F}}
\newcommand{\N}{{\cal N}}
\newcommand{\Lang}{{\cal L}}
\newcommand{\powerset}{{\cal P}}
\newcommand{\FTA}{{\cal FT\!A}}
\newcommand{\Term}{{\sf Term}}
\newcommand{\Empty}{{\sf empty}}
\newcommand{\nonEmpty}{{\sf nonempty}}
\newcommand{\compl}{{\sf complement}}
\newcommand{\args}{{\sf args}}
\newcommand{\preds}{{\sf preds}}
\newcommand{\gnd}{{\sf gnd}}
\newcommand{\lfp}{{\sf lfp}}
\newcommand{\psharp}{P^{\sharp}}
\newcommand{\minimize}{{\sf minimize}}
\newcommand{\headterms}{\mathsf{headterms}}
\newcommand{\solvebody}{\mathsf{solvebody}}
\newcommand{\solve}{\mathsf{solve}}
\newcommand{\fail}{\mathsf{fail}}
\newcommand{\member}{\mathsf{memb}}
\newcommand{\ground}{\mathsf{ground}}

\newcommand{\raf}{{\sf raf}}
\newcommand{\qa}{{\sf qa}}
\newcommand{\spl}{{\sf split}}

\newcommand{\transitions}{\mathsf{transitions}}
\newcommand{\nonempty}{\mathsf{nonempty}}
\newcommand{\dom}{\mathsf{dom}}

\newcommand{\Args}{\mathsf{Args}}
\newcommand{\id}{\mathsf{id}}
\newcommand{\type}{\tau}
\newcommand{\restrict}{\mathsf{restrict}}
\newcommand{\any}{\top}
\newcommand{\dyn}{\top}
\newcommand{\dettypes}{{\sf dettypes}}
\newcommand{\Atom}{{\sf Atom}}

\newcommand{\chc}{{\sf chc}}
\newcommand{\deriv}{{\sf deriv}}

\newcommand{\vars}{\mathsf{vars}}
\newcommand{\Vars}{\mathsf{Vars}}
\newcommand{\range}{\mathsf{range}}
\newcommand{\varpos}{\mathsf{varpos}}
\newcommand{\varid}{\mathsf{varid}}
\newcommand{\argpos}{\mathsf{argpos}}
\newcommand{\elim}{\mathsf{elim}}
\newcommand{\pred}{\mathsf{pred}}
\newcommand{\predfuncs}{\mathsf{predfuncs}}
\newcommand{\project}{\mathsf{project}}
\newcommand{\reduce}{\mathsf{reduce}}
\newcommand{\positions}{\mathsf{positions}}
\newcommand{\contained}{\preceq}
\newcommand{\equivalent}{\cong}
\newcommand{\unify}{{\it unify}}
\newcommand{\Iff}{{\rm iff}}
\newcommand{\Where}{{\rm where}}
\newcommand{\State}{\mathsf{S}}
\newcommand{\qmap}{{\sf qmap}}
\newcommand{\fmap}{{\sf fmap}}
\newcommand{\ftable}{{\sf ftable}}
\newcommand{\Qmap}{{\sf Qmap}}
\newcommand{\states}{{\sf states}}
\newcommand{\head}{\tau}
\newcommand{\atomconstraints}{\mathsf{atomconstraints}}
\newcommand{\thresholds}{\mathsf{thresholds}}
\newcommand{\term}{\mathsf{Term}}
\newcommand{\trees}{\mathsf{trees}}
\newcommand{\renames}{\rho}
\newcommand{\renameps}{\rho_2}
\newcommand{\predicates}{\mathsf{Predicates}}
\newcommand{\query}{\mathsf{q}}
\newcommand{\ans}{\mathsf{a}}
\newcommand{\trace}{\mathsf{tr}}
\newcommand{\constr}{\mathsf{constr}}
\newcommand{\Iproj}{\mathsf{proj}}
\newcommand{\SAT}{\mathsf{SAT}}
\newcommand{\interpolant}{\mathsf{interpolant}}
\newcommand{\unknown}{?}
\newcommand{\rhs}{{\sf rhs}}
\newcommand{\lhs}{{\sf lhs}}
\newcommand{\unfold}{{\sf unfold}}
\newcommand{\arity}{{\sf ar}}
\newcommand{\AND}{{\sf AND}}

\newcommand{\atmost}[1]{\le #1}
\newcommand{\exactly}[1]{=#1}
\newcommand{\exceeds}[1]{>#1}
\newcommand{\anydim}[1]{\ge 0}

\def\ll{[\![}
\def\rr{]\!]}

\newcommand{\sset}[2]{\left\{~#1  \left|
                               \begin{array}{l}#2\end{array}
                          \right.     \right\}}

\newcommand{\qin}{\hspace*{0.15in}}
\newenvironment{SProg}
     {\begin{small}\begin{tt}\begin{tabular}[t]{l}}%
     {\end{tabular}\end{tt}\end{small}}
\def\anno#1{{\ooalign{\hfil\raise.07ex\hbox{\small{\rm #1}}\hfil%
        \crcr\mathhexbox20D}}}

\newtheorem{definition}{Definition}
\newtheorem{example}{Example}
\newtheorem{corollary}{Corollary}

\newtheorem{lemma}{Lemma}
\newtheorem{theorem}{Theorem}
\newtheorem{proposition}{Proposition}

\begin{abstract}

Developing an efficient non-linear Horn clause solver is a challenging task since the solver has to reason about the tree structures rather than the linear ones as in a linear solver.  In this paper we propose an incremental approach to solving a set of non-linear Horn clauses using a linear Horn clause solver. 
%The idea is similar in spirit to  the bounded model checking but the solver only has to deal with a set of linear clauses each time. 
We achieve this by   interleaving a  program transformation and a linear   solver. The program transformation is based on  the notion of \emph{tree dimension}, which we apply  to trees corresponding to Horn clause derivations. 
The dimension of a tree is a measure of its non-linearity -- for example a linear tree  (whose nodes have at most one child) has dimension zero while a complete binary tree has dimension equal to its height. 

A given  set of Horn clauses $P$ can be transformed into a new set of clauses
$P^{\atmost{k}}$ (whose derivation trees are the subset of $P$'s derivation trees with dimension at most $k$).  We start by generating $P^{\atmost{k}}$ with $k=0$, which is linear by definition, then pass it to a linear solver. If $P^{\atmost{k}}$ has a solution $M$, and is a solution to $P$ then $P$ has a solution $M$. If $M$ is not a solution of $P$, we plugged $M$ to $P^{\atmost{k+1}}$ which again becomes linear and pass it to the solver and continue successively for increasing value of $k$ until we find a solution to $P$ or resources are exhausted.   Experiment on some Horn clause verification benchmarks  indicates that this is  a promising approach for solving a set of non-linear Horn clauses using a linear solver. It indicates that many times a solution obtained for some under-approximation $P^{\atmost{k}}$ of $P$ becomes a solution for $P$ for a fairly small value of $k$.

\end{abstract}

\section{Introduction}

In this paper we propose an incremental approach to solving a set of non-linear Horn clauses using a linear Horn clause solver.  The idea is similar in spirit to  the bounded model checking with the following difference. On one hand,   the under-approximation we generate is of bounded dimension (see definition \ref{kdim-most}), which is not finitely nor precisely solvable. On the other hand, each such under-approximation is  linear  and the solver  has to deal with only  linear clauses each time. This is achieved by interleaving a  program transformation and a linear   solver. The program transformation is based on  the notion of \emph{tree dimension}, which we apply  to trees corresponding to Horn clause derivations. 
The dimension of a tree is a measure of its non-linearity -- for example a linear tree  (whose nodes have at most one child) has dimension zero while a complete binary tree has dimension equal to its height. 

We say a set of Horn clause $P$ (including \emph{integrity constraints}) is solvable iff it has a model. \emph{Integrity constraints} are special kind of Horn clauses with  $\false$ head where $\false$ is always interpreted as $\false$.  A given  set of Horn clauses $P$ can be transformed into a new set of clauses
$P^{\atmost{k}}$ (whose derivation trees are the subset of $P$'s derivation trees with dimension at most $k$).  The predicates of $P^{\atmost{k}}$ are subset of predicates of $P$   indexed from $0$ to $k$. We start by generating $P^{\atmost{k}}$ with $k=0$, which is linear by definition, then pass it to a linear solver. If $P^{\atmost{k}}$ has a solution $M$ where $M$ is a set of constrained fact of the form $p_{index}(X) \leftarrow \C(X)$ for each predicate  $p_{index}$ of $P^{\atmost{k}}$. Let  $M'$ is $M$ after removing  \emph{index} of predicates from each constraint fact in $M$. If $M'$ is a solution  to $P$  then $P$ has a solution $M'$. If $M'$ is not a solution of $P$, we plugged $M$ to $P^{\atmost{k+1}}$ which again becomes linear and pass it to the solver and continue successively for increasing value of $k$ until we find a solution to $P$ or resources are exhausted. 

Experiment on some Horn clause verification benchmarks  indicates that this is  a promising approach for solving a set of non-linear Horn clauses. The program transformation allows one to use a linear solver and solution from a previous iteration can be reused in the latter one. Many times a solution obtained for some under-approximation $P^{\atmost{k}}$ becomes inductive, that is, a solution for $P$.

To motivate  readers, we present an example set of constrained Horn clauses (CHCs) $P$ in Figure  \ref{exprogram} which defines the Fibonacci function. This is an interesting  problem whose dimension  depends on the input number  and its computations are trees rather than  linear sequences.
The main contributions of this paper are the following.
 \begin{figure}[t]
\centering
\begin{BVerbatim}
c1. fib(A, B):- A>=0,  A=<1, B=A.
c2. fib(A, B) :- A > 1, A2 = A - 2, fib(A2, B2),
           A1 = A - 1, fib(A1, B1), B = B1 + B2.
c3. false:- A>5, fib(A,B), B<A.          
\end{BVerbatim}
\caption{Example CHCs Fib:  it defines a Fibonacci function.}
%number B  of a number A and contains an integrity constraint relating the input A with the output B.}
 \label{exprogram}

\end{figure}

\begin{enumerate}

\item We give an iterative  algorithm for solving a set of non-linear Horn clauses using a linear solver (Section \ref{procverfication});

\item  We demonstrate the feasibility of our approach in practice  applying it to non-linear Horn clause verification problems (Section \ref{experiments});

\end{enumerate}

\section{Preliminaries}
\label{prelim}

A constrained Horn clause  is a first order  formula of the form 
$ p(X) \leftarrow   \C , p_1(X_1) , \ldots , p_k(X_k) $ ($k \ge 0$) (using Constraint Logic Programming (CLP) syntax),  where $\C$ is a conjunction of constraints with respect to some background theory, $X_i, X$  are (possibly empty) vectors of distinct variables, $p_1,\ldots,p_k, p$ are predicate symbols, $p(X)$ is the head of the clause and $\C , p_1(X_1) , \ldots , p_k(X_k)$ is the body.  A clause is called non-linear if it contains more than one atom in the body ($k > 1$), otherwise it is called linear.  A set of Horn clauses $P$ is called  linear if $P$ only contains linear clauses, otherwise it is called non-linear.  A set of Horn clauses is sometimes called a program. 

\begin{definition}[Solution of Horn clauses]
Let  $P$ be a set of Horn claues and $M$ a  set of constrained fact  of the form $p(X) \leftarrow \C(X)$, where $p(X)$ is a predicate in $P$ 
%except the special predicate ``false'' 
and $\C(X)$ is an interpretation of $p(X)$ under some constraint theory. Then $M$ is called a solution (model) of $P$ if it satisfies each clause in $P$.
%if replacing each predicate  $p(X)$ in  $P$ by  $\C(X)$ makes each clause in $P$ satisfiable. 
\end{definition}

\begin{definition}[Inductive solution of Horn clauses]
\label{inductive-solution}
Let $P_u$ be an under-approximation of a set of Horn clauses $P$. If a solution $M$ of $P_u$  is also a solution of $P$, then $M$ is called an inductive solution of $P$. In other words, a solution of a particular case (that is an under-approximation) of $P$ becomes a solution of $P$.

\end{definition}

\begin{figure}[h!]
  \label{treedimension}
  \centering
    \includegraphics[width=0.70 \textwidth, height=45mm]{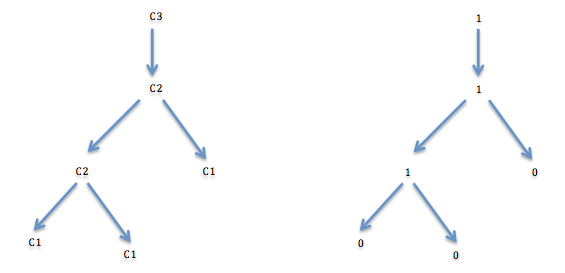}
    \caption{ (a) derivation tree of Fibonacci 3  and (b) its tree dimension.}
\end{figure}

A   labeled tree $c(t_1,\ldots,t_k)$  is a tree with its nodes labeled, where $c$ is a node label and $t_1,\ldots,t_k$ are labeled trees rooted at the children of the node and leaf nodes are denoted by $c$. 
\begin{definition}[Tree dimension (adapted from \cite{DBLP:conf/stacs/EsparzaKL07})]\label{treedim}
  Given a  labeled tree $t= c(t_1,\ldots,t_k)$, the tree dimension of $t$ represented as \(\mathit{dim}(t)\) is defined as follows: 

  \[
  \mathit{dim}(t)= \begin{cases}
  0 & \text{if } k=0 \\ 
  \max_{ i \in [1..k]} \mathit{dim}(t_i) &\text{if  there  is  a  unique  maximum}\\ 
  \max_{ i \in [1..k]} \mathit{dim}(t_i)+1 &\text{otherwise } 
\end{cases}
  \]

\end{definition}
Figure~\ref{treedimension} (a) shows a  derivation tree $t$ for  Fibonacci number  3 and  Figure~\ref{treedimension} (b) shows  its tree dimension.  It can be seen that $\mathit{dim}(t)=1$. This number is a measure of its non-linearity, the smaller the number the closer the tree is to a list.  Since it is not a perfect binary tree, the height of $t$ (3) is greater than its dimension.

 To make the paper self contained, we describe our program transformation approach. Given a set of CHCs $P$ and  $k \in \mathbb{N}$, we split each predicate $H$ occurring in $P$ into the predicates $H^ {\atmost{d}}$ and $ H^ {\exactly{d}}$ where $d \in \{ 0,1,\ldots,k\}$. Here $H^ {\atmost{d}}$ and $ H^ {\exactly{d}}$ generate trees of dimension at most $d$ and exactly $d$ respectively. 

\begin{definition}[ At-most-k-dimension program $P^{\atmost{k}}$]
\label{kdim-most}
 It consists of the following clauses (adapted  from \cite{DBLP:journals/corr/abs-1112-2864, KafleGG2015}):

\begin{enumerate}

\item Linear clauses:

 If $H  \leftarrow  \C \in P$ , then $H^ {\exactly{0}} \leftarrow  \C \in P^{\atmost{k}}$.
 
 If $H  \leftarrow  \C, B_1  \in P$  then $H^ {\exactly{d}} \leftarrow   \C,  B_1 ^{\exactly{d}} \in P^{\atmost{k}}$ for  $0 \le d \le k$.

\item Non-linear clauses: 

 If $H  \leftarrow  \C, B_1 , B_2 , \ldots,B_r  \in P$ with $r>1$:
\begin{itemize}
\item For $1 \le d \le k$, and $1 \le j \le r$:

Set  $Z_j =B_j^{\exactly{d}}$ and $Z_i = B_i ^{\atmost{d-1}}$ for $1 \le i \le r \wedge i \neq j$. Then: 
$H^ {\exactly{d}} \leftarrow   \C, Z_1,\ldots, Z_r \in P^{\atmost{k}}$.

\item 
 For  $1 \le d \le k$, and $J \subseteq \{1,\ldots, r\}$ with $\vert J \vert = 2$:
 
Set $Z_i =B_i^ {\exactly{d-1}}$ if $i \in J$ and $Z_i =B_i^ {\atmost{d-2}}$ if $i \in \{1,\ldots,r\} \setminus J$. If all $Z_i$ are defined, i.e., $d\geq2$ if 
$r > 2$, then:
$H^ {\exactly{d}} \leftarrow   \C, Z_1, \ldots, Z_{r} \in P^{\atmost{k}}$.

\end{itemize}

\item $\epsilon$-clauses: 

$H^ {\atmost{d}} \leftarrow H^ {\exactly{e}} \in P^{\atmost{k}}$ for $0 \le d \le k$ , and every $0 \le e \le d$.

\end{enumerate}
\end{definition}

 \begin{figure}[t]
 \centering
\begin{BVerbatim}
	%linear clauses
        1. fib(0)(A,A) :- A>=0, A=<1.
        2. false(0) :- A>5, B<A, fib(0)(A,B).
        %epsilon-clauses
        3. false[0] :- false(0).
        4. fib[0](A,B) :- fib(0)(A,B).
\end{BVerbatim}
\caption{Fib$^ {\atmost{0}}:$ at-most-0-dimension program of {Fib}.}
 \label{fib0dim}
\end{figure}

The  at-most-0-dimension program  of  {Fib} in Figure \ref{exprogram} is depicted in Figure \ref{fib0dim}.  In textual form we represent a predicate $p^{\atmost{k}}$ by \texttt{p[k]} and a predicate $p^{\exactly{k}}$ by \texttt{p(k)}.
Since some programs have derivation trees of unbounded dimension, trying to verify a property for its increasing dimension separately is not a practical strategy. It only becomes a viable approach  if a solution of $p^{\atmost{k}}$ for some $k\geq 0$ is inductive for $P$ (definition \ref{inductive-solution}).

\section{Solving non-linear set of Horn clauses}
\label{procverfication}
A procedure for solving a set of non-linear Horn clauses using a linear solver is presented in Algorithm \ref{alg:verify}. The main procedure \texttt{SOLVE(P)}  takes a set of non-linear Horn clause as input and outputs a \emph{solution} if they are solvable, \emph{unknown} otherwise. A solution to a set of Horn clause $P$ is an interpretation for the predicates in $P$ that satisfies each clauses in $P$. The procedure makes use of several sub-procedures which will be descirbed next.

\texttt{KDIM(P,k)}  produces an at most $k$ dimesional program $P^{\atmost{k}}$ (ref. definition \ref{kdim-most}). By defintion, $P^{\atmost{k}}$ is linear for $k=0$. For our example program presented in \ref{exprogram}, the at-most-0-dimension program is shown in Figure \ref{fib0dim} and is linear since there is at most one non-constraint atom in the body of each clauses.
\texttt{SOLVE\_LINEAR(P)}  is an oracle (a linear Horn clause solver) that returns a \emph{solution} if $P$ is solvable, otherwise \emph{unsolvable}.  The oracle is sound, that is,  if \texttt{SOLVE\_LINEAR(P)} returns a \emph{solution $S$}  then $P$ is solvable and $S$ is a solution of $P$, if it returns \emph{unsolvable} then $P$ is  unsolvable.
In this paper, we make use of a  solver, which is based on abstract interpretation \cite{DBLP:conf/popl/CousotC77} over the domain of convex polyhedra \cite{Cousot-Halbwachs-78}. The solver produces an over-approximation of minimal model of $P$ as a set of constraints facts for each predictes in $P$. In essence, any Horn clause solver can be used. This approximation is regarded as a solution of the claues if there is no constrained fact for $\false$  in it. Using this solver, the following solution is obtained for the program in Figure \ref{fib0dim}. 

\begin{verbatim}
fib(0)(A,B) :- [-A>= -1,A>=0,B=1].
fib[0](A,B) :- [-A>= -1,A>=0,B=1].
\end{verbatim}

\texttt{INDUCTIVE(M, P)} checks if a given solution $M$  is a solution of $P$. The predicates of $P^{\atmost{k}}$ are subset of predicates of $P$   indexed from $0$ to $k$. If $P^{\atmost{k}}$ has a solution $M$ where $M$ is a set of constrained fact of the form $p_{index}(X) \leftarrow \C(X)$ for each predicate  $p_{index}$ of $P^{\atmost{k}}$. Let  $M'$ is $M$ after removing  \emph{index} of predicates from each constraint fact in $M$. If $M'$ is a solution  to $P$  then $P$ has a solution $M'$.  If this is the case, then \texttt{INDUCTIVE(M, P)} returns ``yes'' otherwise ``no''. Given the above solution  for at-most-0-dimension program, we check if the following ($M'$)
\begin{verbatim}
fib(A,B) :- [-A>= -1,A>=0,B=1].
fib(A,B) :- [-A>= -1,A>=0,B=1].
\end{verbatim}
is a solution to the program in Figure \ref{exprogram}. We can see that it is not the case, since the clause c2 is not satisfied under this solution. It should be noted that after removing the \emph{index} from $M$ to map the solution to the original program, the resulting solution may be in disjunctive form, which allows  a non-disjuntive solver to produce a disjunctive solution. Thanks are  due to our program transformation.

 The sub-procedure \texttt{LINEARIZE(P, M)} generates a linear set of clauses from $P$ and a set of solution $M$. Given $M$ as a set of constrained facts of the form $p(X) \leftarrow \C$, the procedure involves in replacing every predicate $p(X)$ in $P$ with $\C$. This produces a linear set of clauses  under the following condition, which is captured by the following lemma \ref{linear}.

\begin{lemma}[Linearisation]
\label{linear}
If $S$ is a solution computed for a at-most-k-dimension program ($k\geq 0$) and P is a at-most-(k+1)-dimension program, then the procedure  \texttt{LINEARIZE(P, S)} returns a set of linear clauses.
\end{lemma}

An excerpt from a 1-dimension program of our program in Figure \ref{exprogram} is shown below.

\begin{verbatim}
false(1) :- A>5, B<A,  fib(1)(A,B).
fib(1)(A,B) :- A>1, C=A-2, E=A-1, B=F+D,  fib(1)(C,D),  fib[0](E,F).
\end{verbatim}

After plugging in the solution  obtained for at-most-0-dimension program, we obtain the following set of  clauses (after constraints simplification), which are now linear since the body atom \texttt{fib[0](E,F)} in the second clause is replaced by its solution.
\begin{verbatim}
false(1) :-  A>5, B<A, fib(1)(A,B).
fib(1)(A,B) :-  -A>= -2, A>1, A-C=2,  B-D=1, fib(1)(C,D).
\end{verbatim}

Continuing to run our algorithm, the following solution obtained for a 2-dimension program  becomes an inductive solution for the program in Figure \ref{exprogram} and the algorithm terminates. 

\begin{verbatim}

fib(0)(A,B) :- [-A>= -1,A>=0,B=1].
fib[0](A,B) :- [-A>= -1,A>=0,B=1].
fib(1)(A,B) :- [A>=2,A+ -B=0].
fib[1](A,B) :- [A+ -B>= -1,B>=1,-A+B>=0].
fib(2)(A,B) :- [A>=4,-2*A+B>= -3].
fib[2](A,B) :- [A>=0,B>=1,-A+B>=0].

\end{verbatim}

\begin{figure}[t]
\centering
\RestyleAlgo{boxruled}
\LinesNumbered
\begin{algorithm}[H]{Procedure SOLVE ($P$)}

\label{alg:verify}
\caption{Algorithm for solving a set of Horn clauses}
\KwIn{Set of CHCs $P$}
\KwOut{\emph{(solved, solution),  unknown}}
  initialization: $k \gets 0$ \\
   $P' \gets $ KDIM($P,k$) (Definition \ref{kdim-most})\\
  \While {true} { 
   $(r_k, S_k) \gets$ SOLVE\_LINEAR($P'$) \\
  \If {$r_k \neq$ solved} {\Return $unknown$} 
  $b \gets$ INDUCTIVE($S_k, P$) \\
   \If {$b =$ yes} {\Return $(solved, S_k)$} 
    $P_{k+1} \gets $ KDIM($P,k+1$) (Definition \ref{kdim-most})\\
     $ PL_{k+1} \gets$ LINEARIZE($P_{k+1}, S_k$) \\

 $k \gets k+1$ ;  $P' \gets PL_{k+1}$
 
 }
 \vskip 2mm

  \end{algorithm}
  \end{figure}

The soundness of the algorithm \ref{alg:verify} is captured by  the following  \ref{soundness} proposition.
\begin{proposition}[Soundness of algorithm \ref{alg:verify} ] 
\label{soundness}
  If Algorithm \ref{alg:verify} returns a solution $S$ for a set of clauses $P$ then $P$ is solved and $S$ is in fact a solution of $P$.
\end{proposition}

\section{Experimental results}
\label{experiments}
We implemented our algorithm in the tool called \emph{LHornSolver}, which is freely available at \url{https://github.com/bishoksan/LHornSolver}.  Then we carried out an experiment on a set of 9 non-linear CHC verification problems taken from  the repository\footnote{https://svn.sosy-lab.org/software/sv-benchmarks/trunk/clauses/LIA/Eldarica/RECUR/} of software verification benchmarks. Our aim in the current paper is not to make a systematic comparison with other verification techniques;  these are exploratory experiments whether using a linear solver for non-linear problem solving  is practical if at all. The results are summarized in Table~\ref{tbl:exp}. 
%\begin{wraptable}{r}{08cm}
%\caption{Experimental  results on non-linear CHC verification problems}\label{tbl:exp}
%
%    \begin{tabular}{|l|l|l|l|}
%    \hline
%    Program       & Result & Time(ms) & dim(k) \\ \hline
%    addition      & safe   & 4000     & 1      \\ \hline
%    bfprt         & safe   & 2000     & 2      \\ \hline
%    binarysearch  & safe   & 2000     & 1      \\ \hline
%    countZero     & safe   & 2000     & 1      \\ \hline
%    identity      & safe   & 2000     & 1      \\ \hline
%    merge         & safe   & 2000     & 1      \\ \hline
%    palindrome    & safe   & 1000     & 1      \\ \hline
%    fib           & safe   & 2000     & 2      \\ \hline
%    revlen        & safe   & 1000     & 1      \\ \hline
%    avg. time(ms) & ~      & 2000     & ~      \\ \hline
%    \end{tabular}
%
%\end{wraptable}

\begin{table}
\caption{Experimental  results on non-linear CHC verification problems}\label{tbl:exp}
\centering
    \begin{tabular}{|l|l|l|l|}
    \hline
    {\bf Program}       & {\bf Result} & {\bf Time(ms)} & {\bf k} \\ \hline
    addition      & safe   & 4000     & 1      \\ \hline
    bfprt         & safe   & 2000     & 2      \\ \hline
    binarysearch  & safe   & 2000     & 1      \\ \hline
    countZero     & safe   & 2000     & 1      \\ \hline
    identity      & safe   & 2000     & 1      \\ \hline
    merge         & safe   & 2000     & 1      \\ \hline
    palindrome    & safe   & 1000     & 1      \\ \hline
    fib           & safe   & 2000     & 2      \\ \hline
    revlen        & safe   & 1000     & 1      \\ \hline
    avg. time(ms) & ~      & 2000     & ~      \\ \hline
    \end{tabular}

\end{table}
 
In the table {\it Program} represents a program, {\it Result}- represents a verification result using our  approach, {\it Time}-  is a time to solve  the program and {\it k} represents a value  for which a solution of a at-most-k-dimension  program (under-approximation) of a set of clauses $P$ becomes inductive for $P$.

Our  current approach solves 9 out of 9  problems with an average time of 2000 milliseconds. 
%Our previous approach based on proof decomposition   solves these problems with an average time of 4 seconds. 
These examples were also run on QARMC \cite{DBLP:conf/pldi/GrebenshchikovLPR12}  which solves all the problems ( and much faster).

In most of these problems, a solution of an under approximation (at-most-k-dimension program) becomes a solution for the original one when $k=1$. This shows that a solution for an under-approximation (k-dimension program) becomes inductive for a fairly small value of $k$,  and demostrates the feasibility of solving a set of non-linear Horn clauses  using a linear solver.

\section{Related Work}
\label{rel}
There exists several Horn clause solvers \cite{DBLP:conf/tacas/GurfinkelKN15,DBLP:conf/tacas/GrebenshchikovGLPR12,DBLP:conf/cav/RummerHK13,HenriksenGB07,DBLP:conf/vmcai/KafleG15} which handle non-linear clauses. However some tools, for example \cite{DBLP:conf/tacas/AngelisFPP14}, tackle non-linear clauses
by linearising them first and using a linear solver. The generality of the tool such as  \cite{DBLP:conf/tacas/AngelisFPP14} depends on the class of clauses that are linearisable and it is still not clear which classes of clauses are linearisable and under which conditions. But our approach handles any non-linear clauses and solves them using a linear solver.  To the best our knowledge, this is something new.  

\section{Conclusion and future work}

We presented an incremental approach to solving a set of non-linear Horn clauses using a linear Horn clause solver.   We presented an algorithm based on this idea which yielded preliminary results on set of non-linear Horn clause verification benchmarks, showing that the approach is feasible.

\section*{Acknowledgement}
The research leading to these
  results has received funding from the EU 7th
  Framework  318337, ENTRA-Whole-Systems Energy Transparency. The author would like to thank John P. Gallagher and Pierre Ganty for several discussions, suggestions and comments on the earlier version of this paper.

\bibliographystyle{abbrv}
\bibliography{refs}
\end{document}